# Template masks for 4D-STEM

Yining Xie[1,a], Eoin Moynihan[1], Marin Alexe[1], Louis Piper[2], Ana Sanchez[1] and Richard Beanland[1]

1) Department of Physics, University of Warwick, Gibbet Hill Road, Coventry CV4 7AL

2) Warwick Manufacturing Group, University of Warwick, Gibbet Hill Road, Coventry CV4 7AL

a) Yining.Xie@warwick.ac.uk

**Abstract**

We present a new analysis method for atomic resolution four-dimensional scanning transmission electron microscopy (4D-STEM, in which a diffraction pattern is collected at each point of a raster scan of a focused electron beam across the specimen).  In 4D-STEM, each measured intensity has a dual character, forming a pixel in a diffraction pattern and, equally, forming a pixel in a STEM image.  Applying a mask to the data to obtain a 'virtual' bright field or dark field image is widely used and understood. However, there is a complementary procedure, in which an image (template) is applied to the data to obtain a mask.  This mask shows the correlation between the data and the template and, when applied to atomic resolution 4D-STEM data produces an image optimised for the template. This allows, for example, imaging of specific atom columns and is a significant improvement over user-defined masks such as virtual annular bright field imaging. We demonstrate the capability of the approach, separately imaging Li and O atom columns in $LiFePO_4$ and O, Pb and Ti across a domain wall in $PbTiO_3$.  These template masks provide a computationally straightforward and general method to probe 4D-STEM data.  They are particularly effective for specimens of moderate thickness where multiple scattering produces strong and specific correlations in diffraction patterns.

## 1. Introduction

In scanning transmission electron microscopy, the relatively recent arrival of fast pixellated detectors has removed restrictions on detector geometry[1]. Although not yet as fast as conventional detectors [2,3], they have high quantum efficiency and it is possible to collect a complete scattering pattern at every position of the electron probe as it scans across the specimen, producing a four-dimensional dataset (i.e. a 2D image for every point in a 2D raster scan), a technique that has come to be known as 4D-STEM,[1] shown in Fig.1(a). In conventional STEM images, formed by scintillator + photomultiplier electron detectors, the electron flux is integrated across the surface of the scintillator to produce a monolithic signal that is proportional to the sum of electrons detected.  The region of the scattering pattern used to form the image is determined by the microscope's camera length and detector geometry.

Conversely, in 4D STEM there is complete freedom to choose which part of the scattering pattern to select and the weighting it should be given. Here, we present a general method (a *template mask*) that optimises this process for atomic resolution images of crystal structure. We use a two-step process, first defining a template with the aid of an initial STEM image, and then using that template to generate a mask, which can be applied to the data to extract specific features of interest. The method allows visualization of specific atom types in a structure and is applicable to specimens of any thickness where atomic resolution imaging is possible.

Perhaps the most common mode of conventional atomic resolution STEM uses an annular detector that collects electrons scattered to relatively high angles, forming an annular dark field (ADF) image, with bright contrast at atom columns that increases with atomic number (commonly called Z-contrast images). The ready interpretability and relative insensitivity of these images to microscope aberrations, a result of the incoherent scattering that dominates at high angles, makes them the first choice for imaging. However, in materials containing atoms of both high and low atomic number the contrast of light atoms may be so weak that they are effectively invisible in ADF-STEM. Other modes are therefore essential for such materials. Electrons that pass through the central hole of the ADF detector can be collected to produce a bright field (BF) image, in which dynamical (multiple scattering) effects boost the contrast of light atoms, although in thick specimens contrast reversals can occur meaning that interpretation becomes less straightforward. Variations on this geometry include annular bright field (ABF)[4], in which the central part of the BF scattering pattern is blocked, and differential phase contrast (DPC) images, where a segmented annular detector collects the outer part of the direct beam and the difference between opposite segments is used as a signal. [5] ABF provides better contrast for light atoms than BF [4], and is therefore often used to study materials containing oxygen [6], lithium [7, 8], and even hydrogen [9, 10].

Several methods have been developed for better resolution and/or sensitivity to light atoms using 4D-STEM. Integrated Centre of Mass (iCoM), or Integrated Differential Phase Contrast (iDPC) calculate the phase shift of the sample transmission function and thus reveal atomic structure for thin specimens.[11] Ptychography has also proven to be very effective in better resolution and improving dose efficiency for very thin samples, especially 2D materials.[1] However, in specimens more than a few monolayers in thickness, multiple scattering introduces additional phase shifts making the image more complex and harder to interpret. Studies to improve performance for thicker specimens have been performed both for iDPC[12] and focused probe ptychography,[13] but still require clean and damage-free specimens with thicknesses less than a few nm[14, 15], that are difficult to prepare in practice. Most recently, multi-slice ptychography [16] overcomes some of these limitations by modelling the change in electron probe as it propagates through the specimen, providing better resolution and Z-contrast to study thicker samples. Nevertheless, iCoM is rather sensitive to focus, while multi-slice ptychography requires significant post-acquisition computation. A straightforward method to obtain high quality interpretable images of thicker specimens in 4D-STEM remains a challenge.

The improved conventional modes for imaging of light atoms, like ABF, rely on detector geometries that are inflexible. One might therefore expect, in principle, an improvement to be obtained in 4D-STEM by weighting pixels in the detector for regions of the scattering pattern that contain the most information and reducing or eliminating the contribution of others with less information and more noise. Nevertheless, the amount of information in 4D data is large, and not straightforward to analyse. Matrix analysis methods and statistical algorithms, which perform well in handling multidimensional data, are required. At lower magnifications, principal component analysis (PCA), and non-negative matrix factorisation have proven effective for e.g. fast grain mapping and classification of polycrystalline materials,[17] and have been combined with machine learning for the characterisation of microstructures [18]. However, these statistical methods are not, in general, applied to atomic resolution data, in part due to the same fundamental problem of interpretability when multiple scattering dominates. Here, we overcome this problem by using an initial approximation of the desired information to define the data processing route, giving an interpretable result that maximises the use of the signals in a 4D-STEM dataset. Specifically, a *template image* containing specific atom positions acts as an initial 'solution vector' along the scan axis for an individual atomic species. Calculating the Pearson correlation coefficient (PCC) of the 4D-STEM data with this solution vector produces an optimal weighted mask, and application of this mask produces STEM images that represent the best statistical correlation for the position of specific atom types. Following this procedure, multiple structure- and chemistry-specific images may be extracted from the data, each using a different linear combination along the diffraction axis. We show that this approach better resolves atomic structure, especially for thick specimens that are unsuitable for ptychography or iDPC.

In this formulation the act of applying a mask on the diffraction pattern to generate an image, or using structural data such as atomic coordinates to generate a mask, are expressed as matrix operations. As part of this methodology, we reduce the dimensionality of the four-dimensional data by vectorising along diffraction and scan dimensions to give a 2D matrix. Sorting these vectors by scattering angle and structural metrics allows direct visualisation of the whole data set as a single matrix image that simplifies, for example, choice of the optimal ABF signal, by comparing the matrix image with simulation. We investigate the way this approach can be used to enhance contrast of atom columns with low atomic number (i.e. Li and O) in $LiFePO_4$ nanoparticles, widely used in lithium battery cathodes in electric vehicles[19]. As a second example we examine a structure with spatial variations, i.e. a 90° domain wall in $PbTiO_3$, here using a subset of the data to define a mask that is subsequently applied to the whole data. The technique is straightforward both computationally and experimentally and shows promising results on thick (tens of nm) specimens and noisy data.

## 2. Methodology example: optimised ADF-STEM images

We first consider the problem of examining a 4D dataset with a dual nature and begin by noting that each pixel in the diffraction pattern can, ideally, be considered as an independent STEM detector. We call the image produced by a single pixel over the 2D spatial scan, with dimensions $[x, y]$, a *pixel-STEM image*. For a pixellated array with dimensions $[i, j]$, the detector thus produces $i \times j = n$ distinct pixel-STEM images. At the same time, each probe position has an associated diffraction pattern, and there are $x \times y = m$ diffraction patterns. Thus, a measured intensity is both dependent on the position of the pixel in the detector, the position of the probe on the sample, and is strongly correlated with neighbouring pixels (Fig. 1a). This conceptually difficult aspect of 4D-STEM data can be somewhat resolved by reducing its dimensionality. Thus, we convert each diffraction pattern into a 1D row vector **r** of length $n$. The simplest way to do this is by concatenation, i.e. placing the second row in the image into the 1D array following the first, and so on. Doing this for all $m$ diffraction patterns and arranging them as rows gives a 2D $m \times n$ matrix **D**, as shown in Fig. 1b for our [001] LiFePO$_4$ data. In **D** each measured intensity is present only once, and this matrix may be described equally as an array of row vectors **r** (diffraction patterns) or column vectors **s** (pixel-STEM images), i.e.

$$\mathbf{D} = [\mathbf{s}_1, \mathbf{s}_2, \mathbf{s}_3, \cdots, \mathbf{s}_n,] = \begin{bmatrix} \mathbf{r}_1 \\ \mathbf{r}_2 \\ \mathbf{r}_3 \\ \vdots \\ \mathbf{r}_m \end{bmatrix} \qquad (1)$$

While this clarifies the role of each measured intensity, it is not immediately helpful without applying a more useful mapping of the 2D data (images and diffraction patterns) into 1D vectors (columns and rows). In Fig. 1(b), pixel-STEM images from the bright field disk appear as separate vertical stripes because a simple 1D concatenation destroys the 2D spatial correlation present in the original diffraction patterns. A more useful mapping arranges the columns according to the radial distance from the centre of the BF disk and, for atomic resolution data, a further improvement in the visualisation of data in the matrix **D** can be obtained by sorting rows according to their intensity at high scattering angles, as shown in Fig. 1(c). Here, pixels at the centre of the diffraction pattern now form an obvious band on the left, while higher angle scattering is to the right. This organisation of the data allows the information it contains to be more readily visualised.

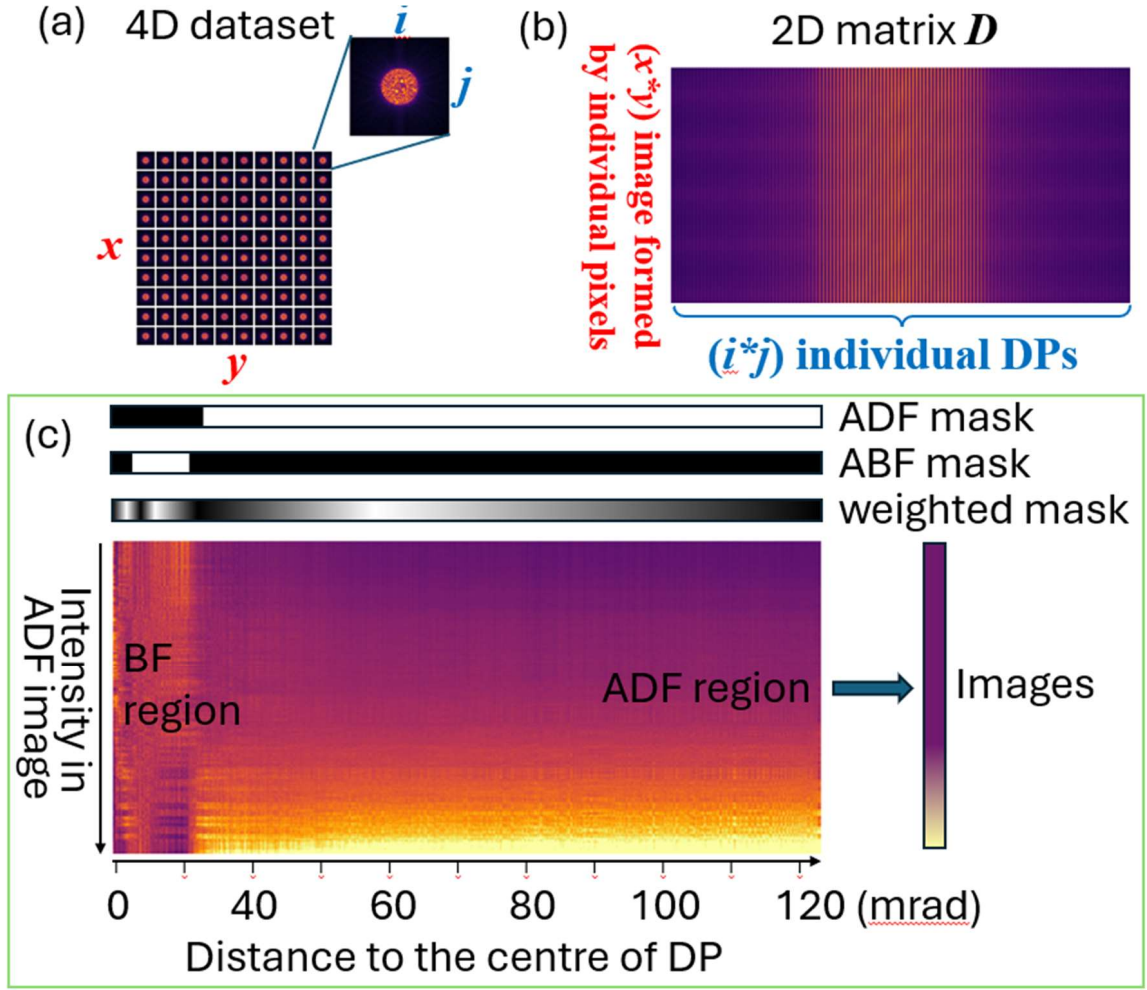

Figure 1 (a) A simulated [001] LiFePO$_4$ 4D-STEM dataset visualised as a $[x, y]$ set of diffraction patterns, each of size $[i, j]$. (b) reduction from 4D to 2D by converting each diffraction pattern to a row vector and stacking them to form a 2D matrix $\mathbf{D}$. Each column is a reshaped pixel-STEM image, corresponding to an individual pixel in the detector. (c) Reorganisation and normalisation of $\hat{\mathbf{D}}$ to visualise the information present in the data; columns from left to right correspond to pixel-STEM images with increasing radial distance from the centre of the BF disk, while rows top to bottom correspond to increasing intensity at high angles. Each column $\mathbf{s}_k$ is normalised according to Eq. (3).

In this formalism, the effect of applying a mask to the data is given by operating the matrix $\mathbf{D}$ on a *mask vector* $\mathbf{r}_m$ of length $n$. Three examples of radial mask are shown in Fig. 1(c) above $\mathbf{D}$, with black representing 0 and white representing 1. Thus, for example, a virtual ADF image $\mathbf{s}_{adf}$ is produced by a mask vector $\mathbf{r}_{adf}$ with value 0 at low angles and 1 at high angles, i.e.

$$\mathbf{s}_{adf} = \mathbf{D}\mathbf{r}_{adf}^T, \qquad (2)$$

with the resulting 1D vector $\mathbf{s}_{adf}$ then reshaped into a 2D STEM image ($T$ indicates the matrix transpose, as required if $\mathbf{r}_{adf}$ is a row vector). Similarly, an ABF mask is described by a vector with value 0 at high and very low angles and value 1 over a range in the BF disk (Fig. 1c).

There is, of course, no reason to restrict these operations to binary masks and our main interest here lies in masks that can have arbitrary weights and shapes, such as the lower mask vector in Fig. 1c. As an introduction to the procedures involved, we first consider ADF-STEM images. Summation of the many pixel-STEM images that comprise the right-hand part of $\mathbf{D}$ produces a virtual ADF image of our atomic resolution $LiFePO_4$ 4D data, Fig. 2a, because the noise in each pixel is uncorrelated and both positive and negative, thus tending to zero when summed, while the signal is always positive, adding to a finite value. As can be seen by comparison with the overlaid crystal structure, in this image only the relatively high atomic number atom columns (Fe and P) are visible, while the oxygen and lithium atom columns give no contrast. (The distortions of the image are due to physical instability of the microscope/electron beam, which is more apparent for 4D-STEM data where the raster scan is relatively slow, limited by the (~2000 Hz) frame rate of the detector.)

The binary ADF mask simply sums all pixel-STEM images outside the centre of the diffraction pattern with equal weights. We seek a more general mask that gives an ADF-STEM image with improved signal-to-noise ratio, which optimises the choice of pixel-STEM images (i.e. the shape of the mask) and how to weight their contributions. We quantify the similarity between each pixel-STEM image and the ADF-STEM image with a metric that is insensitive to differences in magnitude, i.e. the zero-mean normalised cross correlation $r$, also known as the Pearson correlation coefficient, PCC. To calculate the PCC of the ADF-STEM image $\mathbf{s}_{adf}$ with the $k^{th}$ pixel-STEM image in $\mathbf{D}$, the column vector $\mathbf{s}_k$ in $\mathbf{D}$ is normalised by subtracting its mean intensity $\bar{s}_k$ and dividing by its standard deviation $\sigma_{s_k}$, i.e.

$$\hat{\mathbf{s}}_k = \frac{\mathbf{s}_k - \bar{s}_k}{\sigma_{s_k}}, \tag{4}$$

and its PCC in comparison with the ADF-STEM image is then given by

$$r_{adf,k} = \hat{\mathbf{s}}_{adf}^T \cdot \hat{\mathbf{s}}_k, \tag{5}$$

where $\hat{\mathbf{s}}_{adf}$ is a similarly normalised ADF-STEM image. Using Eq. (4) to produce a normalised matrix $\hat{\mathbf{D}}$, we can produce a PCC map for all detector pixels $\mathbf{r}_{adf}$, (a row vector, converted into a 2D image using the same mapping of Eq. 1) which gives the similarity between each pixel-STEM image and the ADF-STEM image:

$$\mathbf{r}_{adf} = \hat{\mathbf{s}}_{adf}^T \hat{\mathbf{D}}. \tag{6}$$

This PCC map reveals the inner structure of the 4D-STEM data, shown for the $LiFePO_4$ 4D dataset in Fig. 2(b). Here a positive correlation is orange, and a negative correlation is blue. The outer part of this PCC map is all positively correlated with the ADF-STEM image, although it is not entirely uniform; it is higher along Kikuchi bands where channelling effects are stronger. The central part of the PCC map, corresponding to the bright field disk of the diffraction pattern, has a complex and symmetrical pattern of positive and negative correlations.

A PCC map $\mathbf{r}$ has the same dimensions as a diffraction pattern and therefore can also be used as a mask that can be applied to the 4D-STEM data, following Eq. (2). Most importantly, in combination with Eq. (6) it is apparent that there is an inherent duality in the equations, which corresponds with the dual nature of measured intensities. That is, application of a mask $\mathbf{r}$ on $\widehat{\mathbf{D}}$ produces a STEM image, $\mathbf{s} = \widehat{\mathbf{D}}\mathbf{r}^T$; and application of an image $\mathbf{s}$ to $\widehat{\mathbf{D}}$ produces a PCC mask $\mathbf{r} = \hat{\mathbf{s}}^T\widehat{\mathbf{D}}$. This complementarity provides a framework with which to investigate 4D-STEM data; any image $\mathbf{s}_{in}$ with dimensions $[x, y]$ can be used to make a mask $\mathbf{r}$, which can then be applied to the 4D-STEM data to produce an image $\mathbf{s}_{out}$

$$\mathbf{s}_{out} = \widehat{\mathbf{D}}\mathbf{r}^T = \widehat{\mathbf{D}}(\hat{\mathbf{s}}_{in}^T\widehat{\mathbf{D}})^T = \widehat{\mathbf{D}}\widehat{\mathbf{D}}^T\hat{\mathbf{s}}_{in} = \mathbf{C}\hat{\mathbf{s}}_{in}, \tag{7}$$

where $\widehat{\mathbf{D}}\widehat{\mathbf{D}}^T = \mathbf{C}$ is the correlation matrix. In the current example, using the PCC map $\mathbf{r}_{adf}$ as a mask produces a STEM image that resembles the input virtual ADF-STEM image, but with significantly higher intensities since all pixel-STEM images contribute to the output (Fig. SI1). Those pixel-STEM images which most resemble the input ADF-STEM image are weighted more strongly, while pixel-STEM images that are uncorrelated with the input are weighted less. Nevertheless, the output image Fig. SI1 has additional contrast in the form of bright horizontal bands, which are unrelated to any atomic structure and are produced by contributions from uncorrelated pixel-STEM images. The similarity between input and output indicates that the ADF-STEM image is almost an eigenvector (principal component) of $\mathbf{C}$ (in which case $\mathbf{s}_{out} = k\mathbf{s}_{in}$, where $k$ is a constant, which can be obtained by eigen decomposition of $\mathbf{C}$, or iterative application of Eq.7). Nevertheless, since our main interest lies in visualising atomic columns the bright bands of contrast in the eigenvector image are not useful and illustrate the limitations of principal component analysis (PCA, also see discussion section below).

There are two factors that give a poor correlation of an atomic resolution pixel STEM image with an input 'test' image (in this case, the virtual ADF-STEM image): first, the information in that pixel may primarily correspond to a different signal, with minima and/or maxima in different positions to that in the test image; and second, the signal may be the same as that in the test image, but the correlation is degraded by noise. Thus, as shown in Fig. 2(c), while pixel-STEM images obtained from >25 mrad from the centre of the PCC map in Fig 2(b) have a positive correlation with the ADF STEM image, their correlation coefficient is low ($r < 0.2$) due to their low intensity and correspondingly high shot noise. Conversely, pixel-STEM images obtained from within the BF disk have high intensity and relatively low noise, but the large range of correlations indicates a variety of images, most of which do not correspond to the desired output. We therefore use a modified scaling in which strong uncorrelated signals are given less weight than weaker correlated, but noisy, signals, i.e. we divide each pixel-STEM image by its mean intensity

$$\hat{\mathbf{s}}'_k = \frac{\mathbf{s}_k - \bar{s}_k}{\bar{s}_k \sigma_{s_k}}, \tag{8}$$

Giving a modified PCC map $\mathbf{r}_{adf} = \hat{\mathbf{s}}_{adf}^T \widehat{\mathbf{D}}'$. An optimised virtual ADF-STEM image Fig. 2(d) is generated from applying this mask to the 4D-STEM data (Eq. 2).

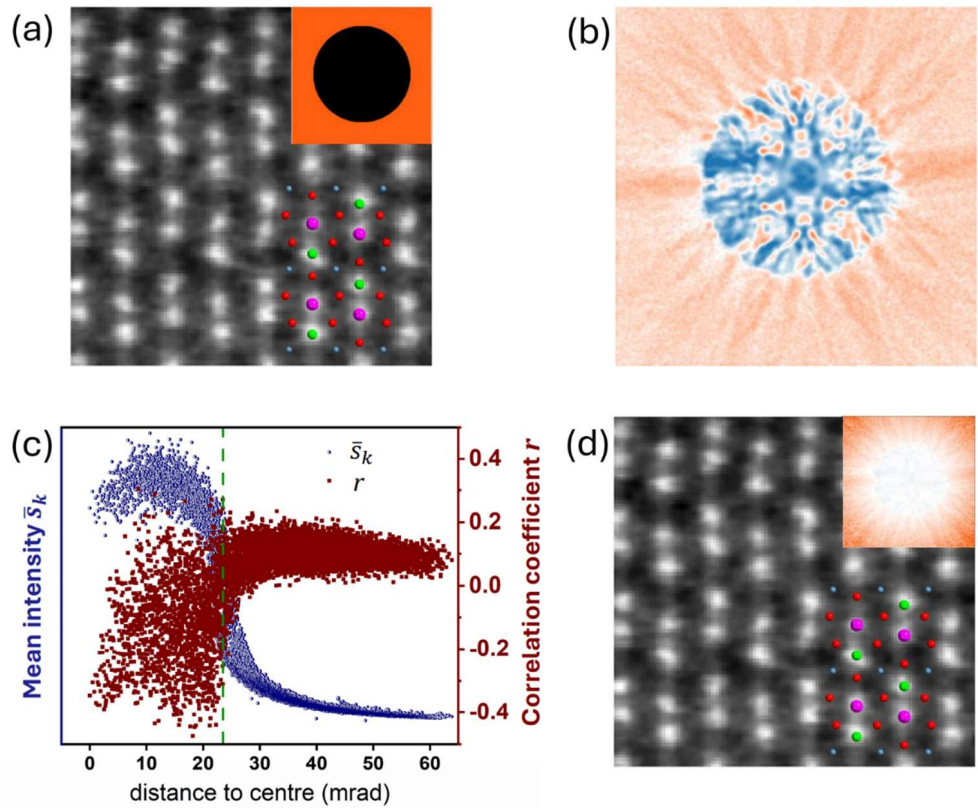

Figure 2. Experimental 4D-STEM data for [001] LiFePO$_4$. (a) virtual ADF-STEM image produced by applying a binary mask to the 4D-STEM data (inset, top right). (b) Map of Pearson correlation coefficient (PCC) $r$ for all pixel-STEM images in comparison with the ADF-STEM image (a). Here, positive correlation is blue and anticorrelation is orange. (c) Scatter plot of pixel-STEM image intensity (blue) and PCC $r$ (red) as a function of scattering angle. (d) optimised ADF-EM image obtained using a normalized PCC mask and a threshold $r > 0$ (inset, top right).

Visually, there is little to distinguish Figs. 2a and 2d, although the latter has an improved signal-to-noise ratio. Both images show 'atomic number' contrast, with Fe ($Z$ = 26) atom columns brighter than the P ($Z$ = 15) atom columns, and the O ($Z$ = 8) and Li ($Z$ = 3) columns essentially invisible. The improvement is marginal in comparison with the output of a conventional ADF detector that integrates signal over a suitable angular range and is essentially negated by the slower acquisition time and the resulting drift. The more important result here is that the mask and weights used to create Fig. 2d do not require a user to select lower/upper scattering angles. As shown below, this approach – using the correlation of pixel-STEM images with a test, or template, image to extract a particular signal of interest – has general applicability. We now turn to the signals that can be uniquely accessed in atomic resolution 4D-STEM, which reside inside the bright field disk.

# 3. Results

## 3.1.  Atom-specific imaging with weighted masks

Annular bright field (ABF) masks are commonly used to improve contrast of light elements, and the image that results from applying a binary ABF mask with inner/outer radii of 11/23.5 mrad to the LiFePO$_4$ 4D-STEM data is shown in Fig. 3a.  Even though the specimen is rather thick – Figs. SI2 and 3 shows that summing all diffraction patterns to give a position-averaged convergent beam electron diffraction (PACBED) pattern and comparison with simulation gives a thickness of ~75 nm – contrast is clearly present in the virtual ABF image that corresponds to the oxygen, and even lithium, atom columns.  In a specimen of this thickness, dynamical diffraction effects dominate contrast mechanisms and there are no obvious guiding principles that can be used to inform the choice of mask.  The choice of inner and outer radii in virtual ABF imaging is therefore rather arbitrary (and is often chosen in practice by varying its dimensions to obtain the 'best' result).  Here, we use the procedures outlined above to generate masks and show that this can be done for each type of atom column in the image.

In the [001] view of LiFePO$_4$ each of the four different atomic species appear in separate and distinct atom columns, as shown in the overlaid structure in Fig. 3(a).  This knowledge, together with a part of the virtual ABF image that allows their approximate locations to be determined, can be used to generate masks that are specific to each atom type, and from these masks images can be extracted from the 4D-STEM dataset as shown in Figs. 3(b)-(e).  Thus, the right part of Fig. 3(a) shows a binary *template image* in which each oxygen atom column is marked by a circular set of pixels.  This oxygen template image $\mathbf{s}_O$ can be used to generate a PCC mask specific to oxygen atom columns (bottom left, Fig. 3b):

$$\mathbf{r}_O = \mathbf{s}_O^T \hat{\mathbf{D}}. \tag{9}$$

This PCC mask shows the regions in the diffraction pattern that have the strongest correlation with the oxygen atom column positions.  (Interestingly, the same mask can be produced using only a fraction of the dataset, allowing masks generated from one measurement to analyse a full experiment.)  As might be expected from the success of ABF in imaging columns of low atomic number,[20, 21] the maximum correlation is found around the zone axis centre, although with a shape that would only be approximated rather poorly by a binary annulus.  This demonstrates how an ABF mask may work, while being clearly sub-optimal for the data.  Applying $\mathbf{r}_O$ to the whole 4D-STEM data following Eq. 7 produces the image bottom left, Fig. 3(c).  Interestingly, only oxygen atom columns are visible and the other low atomic number columns, Li, do not appear.  Performing the same procedure with a Li column template produces a lithium PCC mask $\mathbf{r}_{Li}$ (bottom right, Fig. 3(b)).  Here, the maximum correlation is also found at low angles, although slightly further from the zone axis centre and concentrated in four regions (which could be captured by an annular ABF mask of a suitable size).  However, there are other regions of high correlation at higher angles that could lie outside an ABF mask, as well as regions of low correlation close to the zone axis centre that would not contribute anything useful to an ABF image. Applying $\mathbf{r}_{Li}$ to produce the corresponding image $\mathbf{s}_{Li}$ (bottom right, Fig. 3c) not only reproduces the location of the Li atom columns, but again there is no signal from any other atom column, i.e. the image is chemically specific. The PCC masks of the higher atomic number species, $\mathbf{r}_{Fe}$ and $\mathbf{r}_P$, have regions of both strong positive and negative correlation distributed across the ABF disc. While their corresponding images match the position of their respective atom columns there is some cross talk, i.e. P atoms are faintly visible in the Fe image, and vice-versa, probably a result of the coincidence of regions with strong negative correlations in $\mathbf{r}_{Fe}$ and $\mathbf{r}_P$.

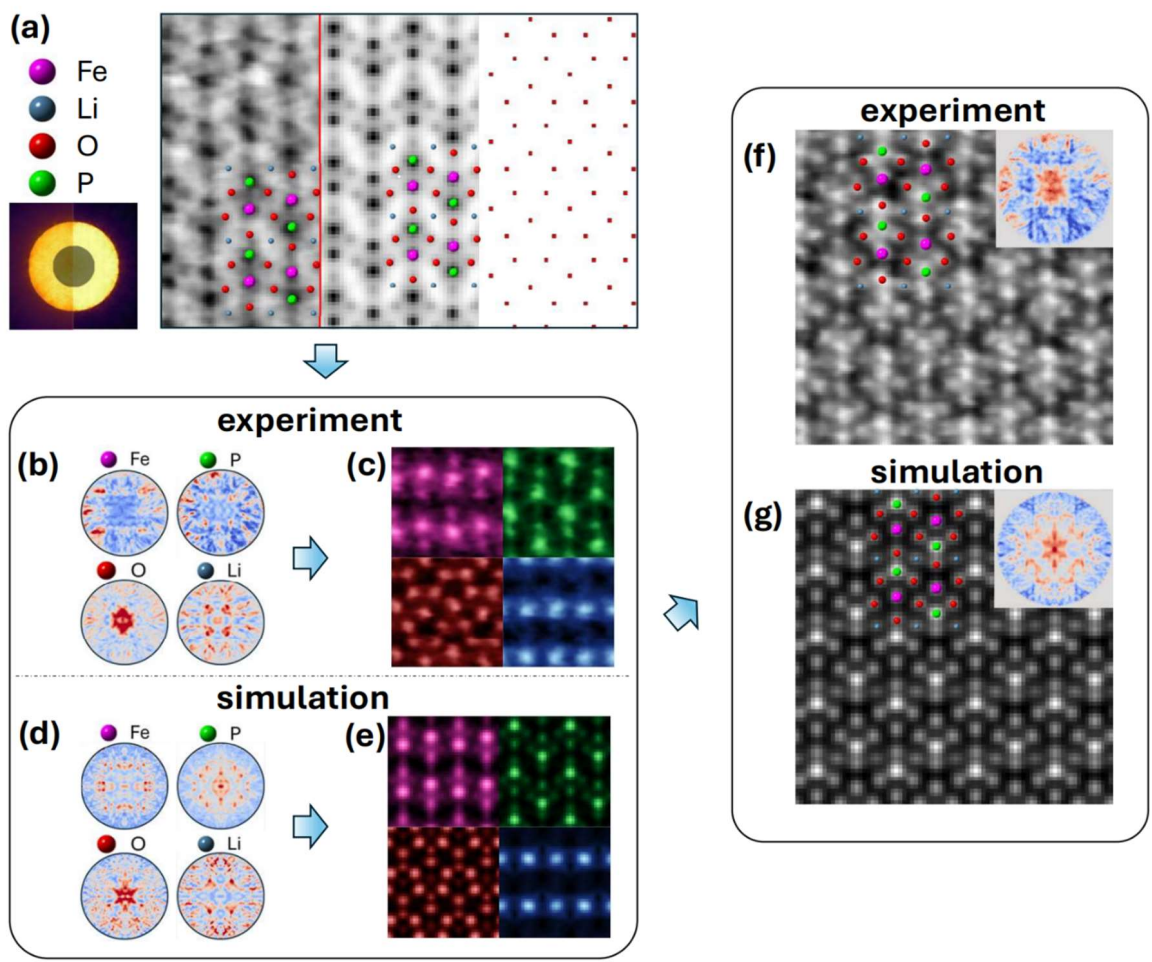

Figure 3. (a) Atom-specific STEM imaging for [001] LiFePO$_4$. (a) ABF image of a 72.5nm thick specimen with corresponding masks (left: experiment, centre: simulation) with overlaid structure showing four types of atom column. A template image marking the position of oxygen columns $\mathbf{s}_O$ is shown on the right. (b) and (d) Atom-specific PCC maps for experimental and simulated data, respectively. (c) and (e) Corresponding atom-specific images obtained by applying the masks in (b) to the 4D-STEM data. (f) and (g) experimental and simulated images for all atom columns (PCC masks inset top right).

Figures 3(d) and 3(e) show a similar set of atom-specific PCC masks and images for a multislice simulated 4D-STEM dataset. The similarity between the PCC masks for simulation and experiment, Figs 3(b) and 3(d), is excellent for O and Li, rather less so for Fe and P. However, the results Figs 3(b) and 3(d) match perfectly, validating the approach and showing that chemically specific template mask imaging is a viable method. It is also possible to produce an optimised image for all atoms using a combined template, as shown in Figs. 2(e, f). These images have significantly better contrast than the ABF image and allow all types of atoms, including Li, to be well resolved.

To summarise our method up to this point: a) we apply a conventional mask (e.g. virtual BF, ABF, or ADF) to the 4D-STEM data set to obtain an initial image; b) sites of interest (e.g. a particular atom type) are chosen in this image; c) a binary template image is constructed, with value zero apart from a set of pixels in these sites, with value 1; d) correlations are determined between the template and all pixel-STEM images, giving a PCC map; e) the PCC map is used as a weighted mask to obtain an output image.

It is not surprising that the final output image is a good match for the input template image, since the procedure is designed to produce exactly this result. However, the specificity of output images to chemically distinct atom columns is both unexpected and potentially very useful. In addition to providing a helpful visualisation of 4D-STEM data, allowing the most sensitive parts of the diffraction pattern for different atom columns to be identified, this approach facilitates the design of weighted masks that will extract chemically specific information in general. We now investigate the capability of the approach for data that contains spatial variations in structure.

### 3.2. Partial templates and microstructure

Our second example is more complex: atomic resolution 4D-STEM data from a 90° ferroelastic/ferroelectric domain wall in $PbTiO_3$, shown in Fig. 4. This perovskite material is tetragonal at room temperature ($c/a$ = 1.06) and has a ferroelectric polarisation along [001] due to the relative displacement of the Ti, O and Pb sublattices. Bulk material generally contains ferroelectric (ferroelastic twin) domains, which form to minimise electrostatic fields. Fig. 4 shows a boundary between two such domains, with an {011} domain wall (DW) running diagonally from top left to bottom right separating regions with the $c$-axis rotated by 90°. As is apparent from the overlaid unit cells, Pb (blue) and Ti (yellow) atom columns are readily visible in both the virtual ADF and ABF images (Figs. 4(a) and 4(b)), but oxygen atom columns (red) give no contrast in the ADF image and cannot be distinguished in the ABF image.

We define *partial templates* by choosing pixels marking the Pb, Ti and the unseen O atom columns using a relatively small reference region, 1/9 of the whole area, (blue box, Fig. 4b). The PCC maps $\mathbf{r}_{Pb}$, $\mathbf{r}_{Ti}$, $\mathbf{r}_O$, and $\mathbf{r}_{all}$ obtained with these templates Figs. 4(c)-(f) are still effective masks for the full data, as is evident from the output images they generate. However, this analysis differs from the first example in an important way. In the case of $LiFePO_4$ atom-specific images could be produced using a binary mask, i.e. simply by setting single pixels at the location of oxygen atom columns to value 1 in an image otherwise of value 0. However, while oxygen columns can be obtained by applying this procedure in this PbTiO$_3$ data, a binary Ti template produces an image in which both Ti and Pb have similar intensities (Fig. SI3)

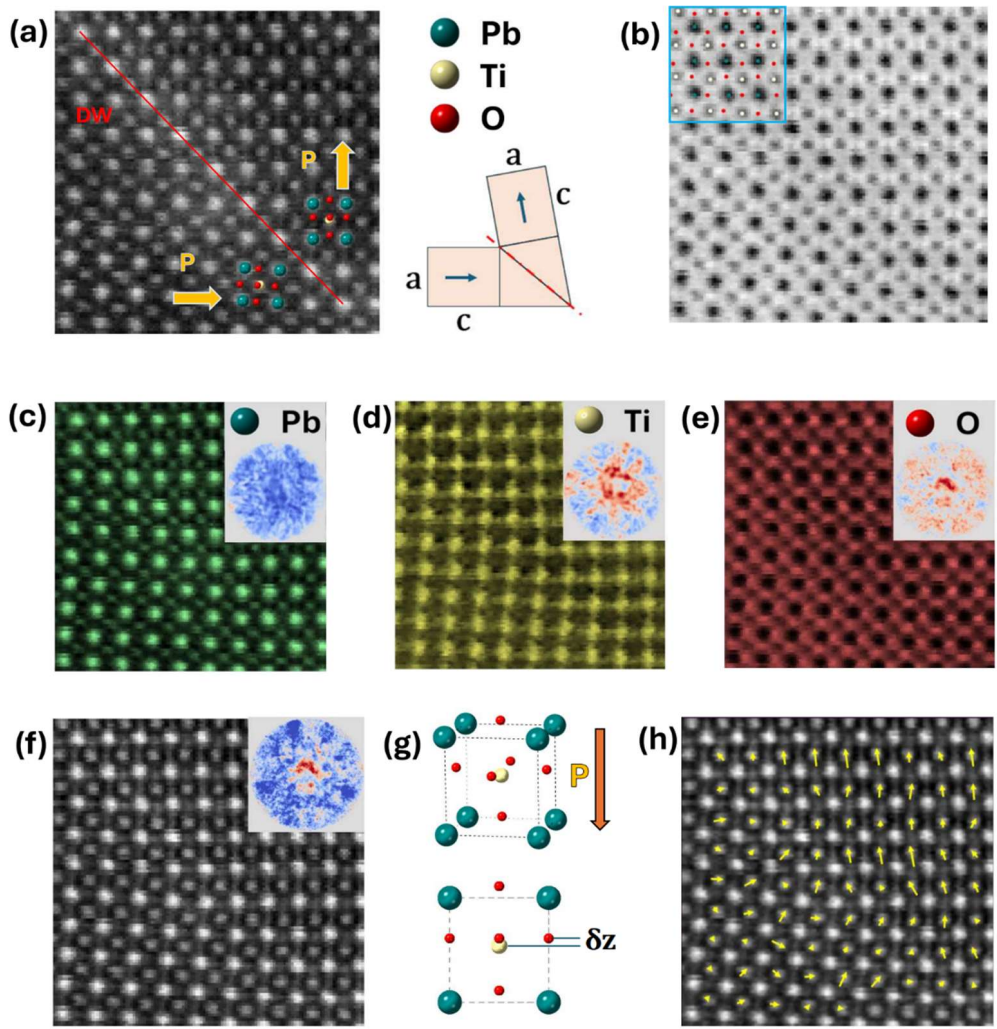

Figure 4. [100] 4D-STEM images of a {011} 90° domain wall in PbTiO$_3$. (a) virtual ADF and (b) virtual ABF images show Pb and Ti+O atom columns, as shown by the overlaid unit cells (Pb = blue, Ti = yellow and O = red). (c), (d) and (e) atom-specific images and (f) all-atom image, each produced by the inset PCC map. (g) atom model of polarized PbTiO$_3$. (h) map of Ti atom column displacement from the centre of its 4 neighbouring oxygen atom columns.

This loss of specificity arises because a binary mask with only a few non-zero pixels works primarily by anti-correlation, i.e. the PCC mask is mainly negative, weighted against pixel-STEM images with poor correlation. In the LiFePO$_4$ data these anticorrelations work well to select specific atom columns, but in the PbTiO$_3$ data they coincide for Ti and Pb. We overcome this problem by using templates that more strongly weight positive correlations, i.e. by selecting a larger number of pixels around the atom column location. Here a 2D gaussian function is used and the balance between positive and negative correlation is tuned by altering its width.

Usefully, even though oxygen atom columns are not visible in either the virtual ADF or ABF images, a partial template using their positions generates an image that is specific to these atom columns. This demonstrates that signals 'hidden' in 4D-STEM data may be accessed using an appropriate template and mask. The polarisation of the material can be determined at a unit cell level by measuring the relative displacements of action and anion sublattices (Fig. 4g). The atom-specific maps allow the relative displacements of Ti and oxygen atom columns to be extracted across the image, showing the change in polarisation at the domain wall (Fig. 4f). The results are in agreement with the previous study of [22].

## 4. Discussion

Several other studies have also considered optimised masks in 4D-STEM imaging. Ahmed et al. [23] used an annular mask for bright field imaging with a negative central region and positive outer region, showing that this approach gave higher contrast for light elements in $LiNiO_2$. Gonnissen et al. examined simulated 4D-STEM data for thin (< 5 nm) [110] $LiV_2O_4$ and (< 30 nm) [001] $SrTiO_3$ [21], with the aim of determining the optimal inner and outer radii of a binary annular mask. They found that that the contrast of atom columns had a strong dependence on the radius selection and was sample dependent. However, neither of these studies considered using the signal of interest to determine the relevant regions in the diffraction pattern, limiting the applicability of their results to the particular materials and specimen thicknesses they investigated. Perhaps the closest approach to our work is that of Krajnak and Etheridge[24], who also used cross correlation to identify symmetrically equivalent points in STEM images. Their approach was based on cross correlations between diffraction patterns after application of symmetry operations, whereas ours examines correlations between images. A combination of the two approaches may be productive.

In contrast our approach, as demonstrated using the above examples, gives a general framework in which template images may be used to extract signals of interest in any atomic resolution 4D-STEM data. There are several other methods commonly applied to these types of data and we now discuss the relationship between them and our approach, in particular principal component analysis (PCA), which has many similarities.

Applying the PCA methodology to 4D-STEM data gives output images (*components*) corresponding to the $n$ eigenvectors of the correlation matrix $\mathbf{C} = \widehat{\mathbf{D}}\widehat{\mathbf{D}}^T$, with the $i^{\text{th}}$ output

$$\mathbf{s}_{[i]} = \lambda_i \mathbf{C} \hat{\mathbf{s}}_{[i]}, \tag{10}$$

i.e. the output image is the same as the input image $\mathbf{s}_{[i]}$ multiplied by a constant, the eigenvalue $\lambda_i$. Following Eq. (7) this can also be reframed as the operation of a mask on the normalised matrix, $\mathbf{s}_{[i]} = \widehat{\mathbf{D}} \mathbf{r}_{[i]}^T$ and we may say that using the input image to create a mask, and using that mask to create an image, simply results in the same image multiplied by a constant. While there is no constraint on the outputs of a principal component analysis that requires them to be physically meaningful, often the largest eigenvalues can be associated with useful parameters. In the case of atomic resolution STEM the largest eigenvalue often has a strong resemblance to an ADF-STEM image, which is clearly the case for the example in section 2, where the virtual ADF-STEM image input produces a mask rather similar to an ADF detector, and the output that results from applying that mask is again an ADF-STEM image. Nevertheless, in PCA it is often unclear whether other eigenvectors have a physical meaning and while in some cases the addition of further constraints can be helpful – such as the requirement that outputs do not have negative values, as is used in non-negative matrix factorisation (NMF) – this is of little use for atomic resolution STEM. In our approach we avoid this problem by choosing a template image with a known correspondence to the structure, and examination of both the PCC mask and the output image aids interpretation of both the diffraction and the image data.

A rather different approach to the analysis of 4D-STEM data can be found in the integrated centre of mass (iCoM) algorithm, where the gradient of the phase of the specimen transmission function is calculate from CoM of the diffraction pattern[11]. This often provides an output image that shows all atom columns in an image but is limited to relatively thin samples in which diffracted intensities are not redistributed in the diffraction pattern by dynamical scattering/channelling effects.

Fig. 5 compares STEM images produced by three different methods. Fig. 5(a-c) show the inverted virtual ABF image, iCoM image and optimised (template mask) image respectively from the same $LiFePO_4$ 4D dataset. The intensity redistribution and increased SNR of (c) in comparison with (a) is readily apparent. For example, light elements in the virtual ABF image (a) are just above the noise level, but they have much more intensity in the template mask image (c), in which the intensity of all 4 elements is similar. The iCoM image (b) obtained from the same data is poor, partly because optimal focus for this method is not the same as conventional STEM[25], and the specimen is relatively thick. A unit-cell averaged line profile along the light elements is given in Fig. 5(d), which sums the normalized intensity between the green lines in (a). The iCoM profile contains peaks that do not correspond to atom columns. However, there is a good match between peaks and atom column positions in both the ABF and template mask images, with an obvious increase of the Li signal in the latter. Fig. 5(e) and (f) show the $PbTiO_3$ data. In this thinner specimen the iCoM image Fig. 5(e) is still not under the best focus condition, but all atoms may be resolved. In Fig. 5(f) the template mask image (right) has better SNR and identifiable oxygen columns in comparison with inverted ABF (left) and while the oxygen atom columns have much lower intensity in comparison with iCoM the ability to extract them independently (Fig. 4e) overcomes this issue. Since the signals become more localised in the diffraction pattern for thicker samples, we expect that iCoM is somewhat complementary to our template approach in terms of specimen thickness.

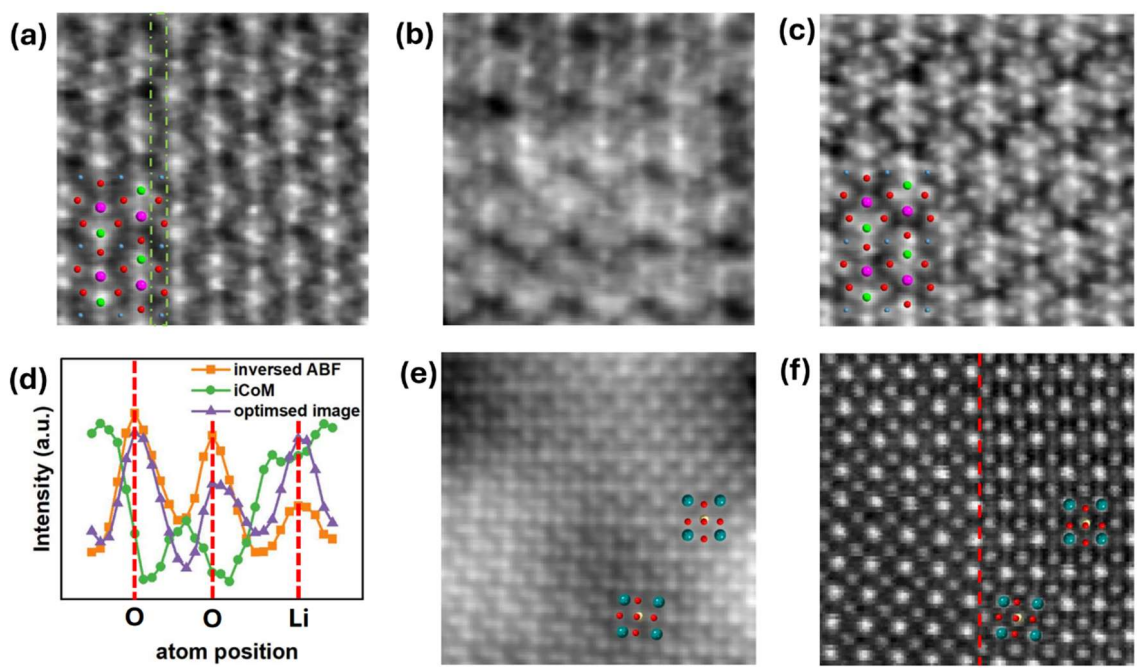

Figure 5. Comparison between different imaging modes. (a-c) LiFePO$_4$, (a) inverted virtual ABF image, (b) iCoM, and (c) template mask image (all atoms). (d) the line profile between the green lines in (a), normalized and averaged over unit cells. (e - f) PbTiO$_3$, (e) iCoM, and (f) inverted ABF (left), template mask image (all atoms, right).

Perhaps the most useful result of the template mask approach is the ability to produce chemically specific images, as shown in Figs. 2 and 3. Sensitivity of electron diffraction data to specific atom types is well documented, for example using the ALCHEMI approach to enhance spectroscopic signals under particular diffraction conditions[26], while Roussow et al.[27] produced energy-filtered electron channelling patterns that showed strong variations in intensity for different atom types. These methods rely on dynamical scattering and channelling to enhance the sensitivity to different atom columns, which is consistent with the observation that template mask imaging works well in specimens of moderate thickness. An obvious next step that builds on these links would be to again return to spectroscopy; for example, in electron energy loss spectroscopy (EELS) choosing regions of the diffraction pattern with strong correlation to particular atom types to pass into the spectrometer should significantly enhance their signal.

## 5. Summary and conclusion

The central idea of our approach is to use the 4D-STEM data itself to generate weighted masks instead of, e.g., a user-generated annular mask. This new method ensures that the signal in the diffraction pattern is used more efficiently. In brief, we generate a Pearson cross-correlation (PCC) map from a template image that selects for particular atom columns. Using this PCC map as a weighted mask then generates STEM images specific to those atom columns. We trial this approach on two examples, a LiFePO$_4$ sample and a domain wall in PbTiO$_3$. Our experimental results appear robust and match well with simulations. In LiFePO$_4$, we obtain strong and specific signals for Li and O atom columns, outperforming the iCoM method in this ~70nm thick sample. In PbTiO$_3$, we show that a template mask calculated from a small homogeneous part of the data may be applied across a spatially inhomogeneous image and still yield useful information (e.g. oxygen atom column locations). Since the template mask method uses a linear combination of the diffraction patterns in the same way as virtual bright or dark field imaging, it is a relatively direct imaging method that is computationally straightforward, giving interpretable images that could be obtained during live imaging.

In conclusion, the template mask approach described in this work provides a 4D STEM method that extracts specific signals of interest from atomic resolution data. These signals include chemically-specific STEM images for particular atom types, including those of low atomic number. It also appears particularly suited to imaging specimens that are too thick for other method such as iCoM or ptychography.

## 6. Experiment

The LiFePO$_4$ TEM specimen was prepared by mixing LiFePO$_4$ nanoparticles with fine aluminium powder (approx. 10:1 Al:LiFePO$_4$) in an Al foil wrap, which was cold rolled to produce a solid Al sheet approx. 100 µm thick. A piece of the sheet was mechanically ground, polished, and ion milled to electron transparency using 6 kV Ar$^+$ ions, with final 1 kV and 0.1 kV cleaning to remove surface damage. The PbTiO$_3$ TEM specimen was prepared from flux growth PbTiO$_3$ single crystals using a Tescan Amber FIB-SEM. The lamella was thinned using Ga ion beam at 30 kV, 50~150pA, and polished using 2~5 kV, 20~100pA.

Experimental 4D-STEM datasets were acquired at room temperature in a double aberration-corrected JEOL ARM200F using a Quantum Detector Merlin pixelated Detector. The datasets were taken at 200 kV accelerating voltage with ~23.5 mrad convergent beam semi-angle. The [001] LiFePO$_4$ dataset was taken from a ~72.5 nm thick region, while the [100] PbTiO$_3$ dataset was taken from a ~44 nm thick region as determined using PACBED, shown in Fig. SI8. py4DSTEM[28] was used to process the data and generate iCoM images.

4D-STEM Simulations were performed using abTEM[29] using conditions to match experiment (Table 1).

| Parameter | Value | Parameter | Value |
|---|---|---|---|
| Thickness | 20~90 nm | Defocus | 0 nm |
| Crystal size | 80 x 80 x 900 Å | Cs | 3 µm |
| Slice thickness | 4.692 Å | Probe soft edge | 4 mrad |
| model | Kirkland | Probe sampling (angular) | 0.334 |
| projection | finite | Energy | 200 kV |
| Potential sampling | 0.073 Å | Detector range | 64.2 mrad |
| Frozen Phonons | 20 iters | Scan step size | ~0.3 Å |

## 7. Acknowledgements

The authors acknowledge Warwick Electron Microscopy RTP for the use of the instruments.

Yining Xie would acknowledge the foundation from China Scholarship Council (CSC).